\documentclass[runningheads]{llncs}

\usepackage{mathptmx} 
\usepackage{microtype} 
\usepackage{paralist}  
\usepackage{mdwlist}

\usepackage[utf8]{inputenc}

\usepackage{multirow} 
\usepackage{graphicx}

\usepackage{listings, xcolor}

\definecolor{verylightgray}{rgb}{.97,.97,.97}

\lstdefinelanguage{Solidity}{
	keywords=[1]{anonymous, assembly, assert, balance, break, call, callcode, case, catch, class, constant, continue, contract, debugger, default, delegatecall, delete, do, else, event, export, external, false, finally, for, function, gas, if, implements, import, in, indexed, instanceof, interface, internal, is, length, library, log0, log1, log2, log3, log4, memory, modifier, new, payable, pragma, private, protected, public, pure, push, require, return, returns, revert, selfdestruct, send, storage, struct, suicide, super, switch, then, this, throw, transfer, true, try, typeof, using, value, view, while, with, addmod, ecrecover, keccak256, mulmod, ripemd160, sha256, sha3}, 
	keywordstyle=[1]\color{blue}\bfseries,
	keywords=[2]{address, bool, byte, bytes, bytes1, bytes2, bytes3, bytes4, bytes5, bytes6, bytes7, bytes8, bytes9, bytes10, bytes11, bytes12, bytes13, bytes14, bytes15, bytes16, bytes17, bytes18, bytes19, bytes20, bytes21, bytes22, bytes23, bytes24, bytes25, bytes26, bytes27, bytes28, bytes29, bytes30, bytes31, bytes32, enum, int, int8, int16, int24, int32, int40, int48, int56, int64, int72, int80, int88, int96, int104, int112, int120, int128, int136, int144, int152, int160, int168, int176, int184, int192, int200, int208, int216, int224, int232, int240, int248, int256, mapping, string, uint, uint8, uint16, uint24, uint32, uint40, uint48, uint56, uint64, uint72, uint80, uint88, uint96, uint104, uint112, uint120, uint128, uint136, uint144, uint152, uint160, uint168, uint176, uint184, uint192, uint200, uint208, uint216, uint224, uint232, uint240, uint248, uint256, var, void, ether, finney, szabo, wei, days, hours, minutes, seconds, weeks, years},	
	keywordstyle=[2]\color{teal}\bfseries,
	keywords=[3]{block, blockhash, coinbase, difficulty, gaslimit, number, timestamp, msg, data, gas, sender, sig, value, now, tx, gasprice, origin},	
	keywordstyle=[3]\color{violet}\bfseries,
	identifierstyle=\color{black},
	sensitive=false,
	comment=[l]{//},
	morecomment=[s]{/*}{*/},
	commentstyle=\color{gray}\ttfamily,
	stringstyle=\color{red}\ttfamily,
	morestring=[b]',
	morestring=[b]"
}

\definecolor{darkblue}{rgb}{0.0, 0.0, 0.55}
\definecolor{darkred}{rgb}{0.55, 0.0, 0.0}
\definecolor{forestgreen}{rgb}{0.13, 0.55, 0.13}

\newcommand{\newtool}{sCompile}

\newcommand{\instr}[1]{{\color{purple} \tt{#1}}}

\usepackage[skip=3pt]{caption}
\newcommand{\newtoolBold}{\textbf{{\textit{sCompile}}}}
\usepackage{wrapfig}
\newenvironment{infobox}{\wrapfigure{l}{.2\textwidth}}{\endwrapfigure}

\usepackage[belowskip=-10pt,aboveskip=0pt]{caption}
 
\begin{document}
\title{\newtool: Critical Path Identification and Analysis for Smart Contracts}

\author{Jialiang Chang\inst{1} \and
Bo Gao\inst{2} \and
Hao Xiao\inst{2} \and
Jun Sun\inst{3} \and
Yan Cai\inst{4} \and
Zijiang Yang\inst{1}}
\authorrunning{Chang et al.}
%
\institute{Department of Computer Science, Western Michigan University, Kalamazoo MI 49009, USA\\
\email{\{jialiang.chang,zijiang.yang\}@wmich.edu} \and
Pillar of Information System Technology and Design, Singapore University of Technology and Design, Singapore\\
\email{bo\_gao@mymail.sutd.edu.sg},\email{hao\_xiao@sutd.edu.sg} \and
School of Information Systems, Singapore Management University, Singapore\\
\email{junsun@smu.edu.sg} \and
State Key Laboratory of Computer Science, Institute of Software, Chinese Academy of Sciences, Beijing, China\\
\email{ycai.mail@gmail.com}
}

\maketitle              
\begin{abstract}
Ethereum smart contracts are an innovation built on top of the blockchain technology, which
provides a platform for automatically executing contracts in an anonymous, distributed, and trusted way. The problem is magnified by the fact that smart contracts, unlike ordinary programs, cannot be patched easily once deployed. It is important for smart contracts to be checked against potential vulnerabilities.

In this work, we propose an alternative approach to automatically identify critical program paths (with multiple function calls including \emph{inter-contract} function calls) in a smart contract, rank the paths according to their criticalness, discard them if they are infeasible or otherwise present them with user friendly warnings for user inspection. We identify paths which involve monetary transaction as critical paths, and prioritize those which potentially violate important
properties. For scalability, symbolic execution techniques are only applied to top ranked critical paths. Our approach has been implemented in a tool called \newtool, which has been applied to 36,099 smart contracts. The experiment results show that \newtool~is efficient, i.e., 5 seconds on average for one smart contract. Furthermore, we show that many known vulnerabilities can be captured if user inspects as few as 10 program paths generated by \newtool. Lastly, \newtool~discovered 224 unknown vulnerabilities with a false positive rate of 15.4\% before user inspection.

\end{abstract}

\section{Introduction}\label{sub:introduction}
Built on top of cryptographic algorithms~\cite{Diffie:2006,Diffie:1976,jorstad1997} and the blockchain technology~\cite{haber1990time,brito2013bitcoin,narayanan2016bitcoin}, cryptocurrency like Bitcoin has been developing rapidly in recent years. Many believe it has the potential to revolutionize the banking industry by allowing monetary transactions. Smart contracts bring it one step further by providing a framework which allows any contract to be executed in an autonomous, distributed, and trusted way. Smart contracts thus may revolutionize many industries. Ethereum~\cite{wood2014ethereum}, an open-source, blockchain-based cryptocurrency, is the first to integrate the functionality of smart contracts. Due to its enormous potential, its market cap reached at \$29.1 billion as of Jun 17th, 2019.

In essence, smart contracts are computer programs which are automatically
executed on a distributed blockchain infrastructure. A majority of smart contracts in Ethereum are written in a programming language called Solidity~\cite{ethereum90:online}. Like ordinary programs, Solidity programs may contain vulnerabilities, which potentially lead to attacks. The problem is magnified by the fact that smart contracts, unlike ordinary programs, cannot be patched easily once they are deployed on the blockchain.

In recent years, these attacks exploit security vulnerabilities in Ethereum smart contracts and often result in monetary loss. One notorious example is the DAO attack~\cite{AttackTh83:online}, i.e., an attacker stole more than 3.5 million Ether (about \$45 million USD at the time) from the DAO contract on June 17, 2016.

The problem of analyzing and verifying smart contracts is far from being
solved. Some believe that it will never be, just as the verification problem of
traditional programs. Solidity is designed to be Turing-complete which intuitively means that it is very expressive and flexible. The price to pay is that almost all interesting problems associated with checking whether a smart contract is vulnerable are undecidable~\cite{turing1937computable}. Consequently, tools which aim to analyze smart contracts \emph{automatically} either are not scalable or produce many false alarms. For instance, Oyente~\cite{luu2016making} is designed to check whether a program path leads to a vulnerability or not using a constraint solver to check whether the path is feasible or not. Due to the limitation of constraint solving techniques, if Oyente is unable to determine whether the path is feasible or not, the choice is either to ignore the path (which may result in a false negative, i.e., a vulnerability is missed) or to report an alarm (which may result in a false alarm).

Besides, we believe that manual inspection is unavoidable given the expressiveness of Solidity. However, given that smart contracts often enclose many behaviors (which manifest through different paths), manually inspecting every path is overwhelming. Thus, \newtool~further aims to reduce the manual effort by identifying a small number of critical paths and presenting them to the user with easy-to-digest information. 

Overall, \newtool~works as follows: 

\begin{compactitem}
  \item \newtool~firstly constructs a control flow graph (CFG) which captures all possible control flow including those due to the \emph{inter-contract} function calls. \newtool~then systematically generates paths (with a bounded sequence of function calls).
  \item To address path explosion, \newtool~then statically identifies paths which are `critical'. In this work, we define paths involving monetary transactions as critical paths, which is often sufficient in capturing vulnerabilities in smart contracts.
  \item  

  We then define a set of (configurable) money-related properties based on existing vulnerabilities and identify all paths that potentially violate our properties. Considering that different properties have different criticalness and a long path may be unlikely feasible than a short one, \newtool~ranks all paths by computing a criticalness score for each path based on the two factors.
  \item Finally, for top ranked paths,
  \newtool~automatically checks whether it is feasible using symbolic execution techniques. 
  And, the feasible paths are presented to the user for inspection.
\end{compactitem}

We have implemented \newtool~and applied it to 36,099 smart contracts gathered from EtherScan~\cite{Ethereum27:online}. Our experiment shows that \newtool~can efficiently analyze smart contracts, i.e., it spends 5 seconds on average to analyze a smart contract (with a bound
on the number of function calls $3$). 
Furthermore, we show that \newtool~effectively prioritizes programs paths which reveal vulnerabilities in smart contracts, i.e., it is often sufficient to capture the vulnerability by inspecting the reported $10$ or fewer critical paths. Overall, \newtool~identified 224 vulnerabilities. The false positive rate of \newtool~(before the results are reported for user inspection) is 15.4\%, which is also generally acceptable. A further user study result shows that with \newtool's help, users are more likely to identify vulnerabilities in smart contracts.

The rest of the paper is organized as follows. Section~\ref{examples} illustrates how \newtool~works through a few simple examples. Section~\ref{approach} presents the details of our approach step-by-step. Section~\ref{experiment} shows evaluation results on \newtool. Section~\ref{related} reviews related work and lastly Section~\ref{conclusion} concludes with a discussion on future work.

\section{Illustrative Examples} \label{examples}
In this section, we present multiple examples to illustrate vulnerabilities in smart contracts and how \newtool~helps to reveal them. The contracts are shown in Fig.~\ref{contract:Illustrative_contracts}. 

\textbf{Example 1:} Contract \emph{toyDAO} is an invariant one of DAO contract. Mapping \emph{credit} is a map which records a user's credit amount. Function \emph{donate()} allows user to top up its credit with $100$ wei (which is a unit of Ether). Function \emph{withdraw()} by design sends $20$ wei to message sender (at line 1) and then updates \emph{credit}. However, when line 1 is executed, message sender could call function \emph{withdraw()} through its fallback function, before line 2 is executed. Line 1 is then executed again and another $20$ wei is sent to message sender. Eventually, all Ether in this contract's wallet is sent to message sender.

In \newtool, inspired by common practice in banking industry, assume that the user sets the limit to be 30. Given the contract, a critical path reported by \newtool~is one which executes line 0, 1, 0, and 1. The path is associated with a warning message stating that the accumulated amount transferred along the path is more than the limit.\textbf{} We remark that existing approaches often check such vulnerability through a property called reentrancy, which often results in false alarms~\cite{luu2016making,kalra2018zeus}.

\begin{figure}[t]
{\scriptsize
  \begin{verbatim}
  contract toyDAO{
      address owner;
      mapping (address => uint) credit;
      function toyDAO() payable public {
          owner = msg.sender;
      }
      function donate() payable public{
          credit[msg.sender] = 100;
      }
      function withdraw() public {
 0        uint256 value = 20;
 1        if (msg.sender.call.value(value)()) {
 2            credit[msg.sender] = credit[msg.sender] - value;
          }
      }
  }
  contract Bitway is ERC20 {
      function () public payable {
          createTokens();
      }
      function createTokens() public payable {
          require(msg.value > 300);
          ...
      }
      ...
  }
  \end{verbatim}}
\vspace{-10pt}
  \caption{Illustrative contracts}
  \vspace{-5pt}
  \label{contract:Illustrative_contracts}
  \vspace{-.2cm}
\end{figure}

\textbf{Example 2:}
Contract \emph{Bitway} is another token management contract. 
It receives Ether (i.e., cryptocurrency in Ethereum) through function \emph{createTokens()}. Note that this is possible because function \emph{createTokens()} is declared as \emph{payable}. However, there is no function in the contract which can send Ether out. Given this contract, \newtool~identifies a list of critical paths for user inspection. The most critical one is a path where function \emph{createTokens()} is invoked. Furthermore, it is labeled with a warning message stating that the smart contract appears to be a ``black hole'' contract as there is no path for sending Ether out, whereas this path allows one to transfer Ether into the wallet of the contract. By inspecting this path and the warning message, the user can capture the vulnerability. In comparison, existing tools like Oyente~\cite{luu2016making} and MAIAN~\cite{nikolic2018finding} report no vulnerability given the contract. We remark that even although MAIAN is designed to check similar vulnerability, it checks whether a contract can receive Ether through testing\footnote{MAIAN sends a value of 256 wei to the contract deployed in the private blockchain network} and thus results in a false negative in this case.

\begin{figure*}[!h]
\includegraphics[width=13cm,height=3cm,scale=0.9]{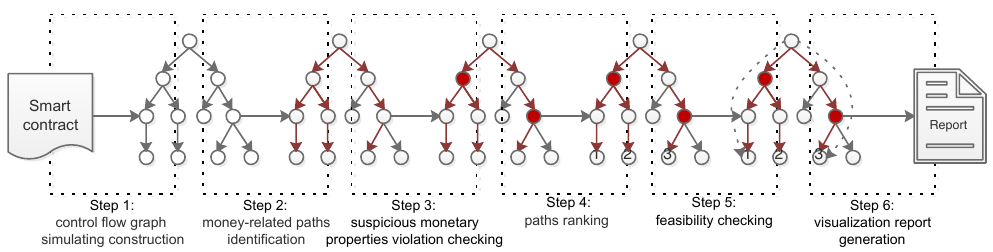}
\vspace{-5pt}
\caption{Overall workflow of \newtool}
\vspace{-10pt}
\label{overallworkflow}
\end{figure*}

\section{Approach} \label{approach}
Fig.~\ref{overallworkflow} shows the overall work flow of \newtool. 
Firstly, given a smart contract, \newtool~constructs a control flow graph (CFG)~\cite{allen1970control} and systematically enumerates all paths. Secondly, we identify the monetary paths based on the CFG up to a user-defined bound on the number of function calls. Thirdly, we analyze each path in order to check whether it potentially violates any of the pre-defined monetary properties. Next, we compute a criticalness score for each and rank the paths accordingly. Afterwards, we apply symbolic execution to filter infeasible critical paths. 
Lastly, we present the results along with the associated paths to the user for inspection.

\subsection{Constructing CFG}
\newtool~constructs a CFG for a smart contract (the compiled EVM opcode with a single entrance for whole and for each function) to capture all possible paths. Formally, a CFG is a tuple $(N, root, E)$ such that
\begin{compactitem}
    \item $N$ is a set of nodes, where each node is a basic block of opcodes.
    \item $root \in N$ is the first basic block of opcodes.
    \item $E \subseteq N \times N$ is a set of edges, where each edge $(n, n')$ corresponds to exactly a control directly from flow $n$ to $n'$.
\end{compactitem}

We also consider inter-contract functions calls, where there is a \instr{CALL} to a foreign function that is assumed to call the current function including third-part contract.

\begin{figure}[t]
\centering
\includegraphics[width=8cm,height=6cm,scale=0.9]{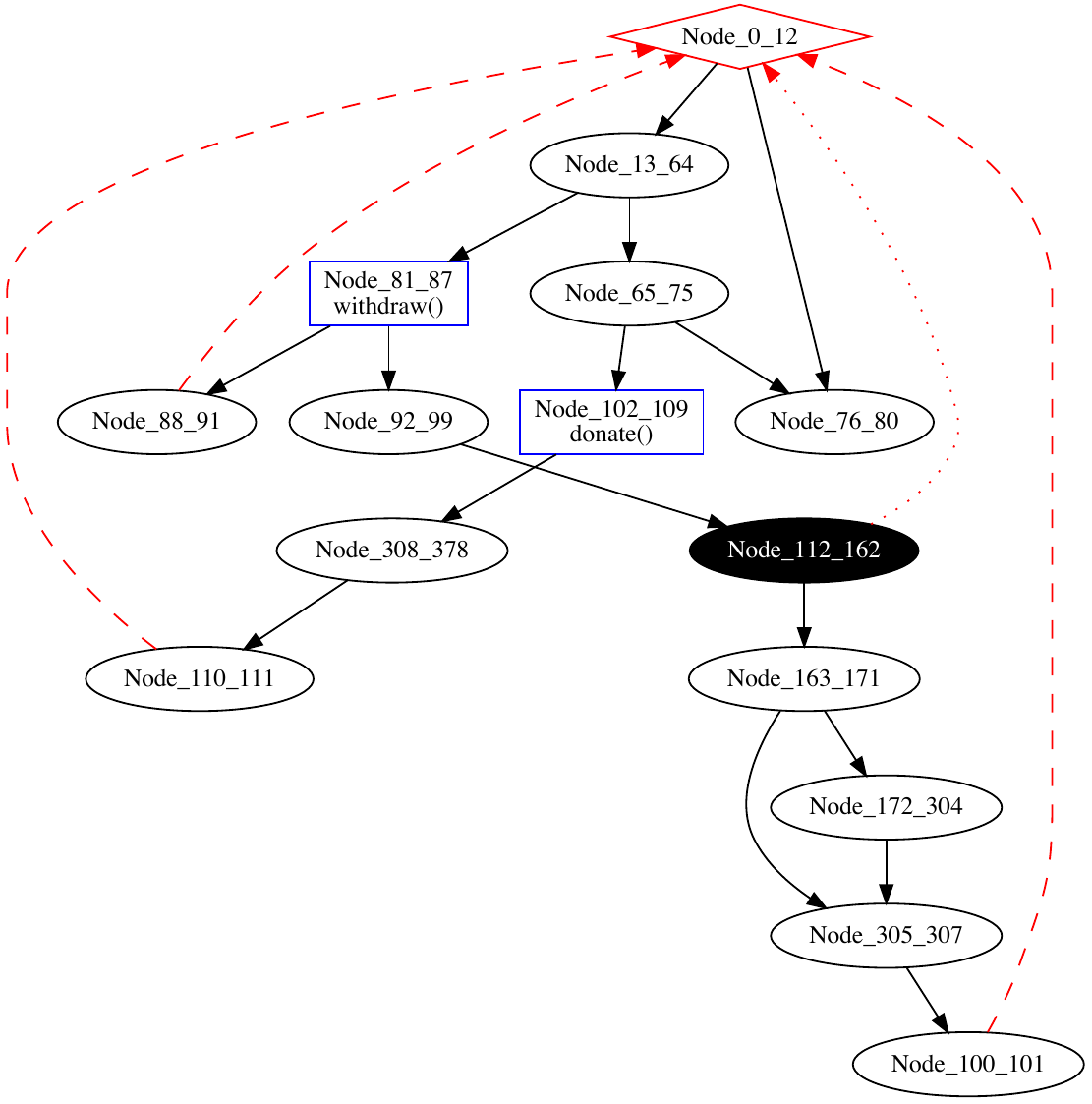}
\caption{Control flow graph of the contract \emph{toyDAO}}
\label{toyDAOCFG}
\vspace{-10pt}
\end{figure}

For instance, Fig.~\ref{toyDAOCFG} shows the CFG of contract
\emph{toyDAO} shown in Fig.~\ref{contract:Illustrative_contracts}. 
Each node is in the form of $Node\_m\_n$, where $m$ and $n$ are indices of the first and the last opcodes of the basic block, respectively. 
The red diamond node at the top is the $root$ node; the blue rectangle nodes represent the first node of a function. 
Note that a black oval represents a node 
that can be redirected to the root due to inter-contract function calls. 
The black solid edges represent the normal control flow. The red dashed edges represent control flow due to a new function call, e.g., the edge from $Node\_88\_91$ to $Node\_0\_12$. That is, for every node $n$ such that $n$ ends with a terminating opcode instruction (i.e., \instr{STOP}, \instr{RETURN}), we introduce an edge from $n$ to $root$. The red dotted edges represent control flow due to the inter-contract function call. That is, for every node which ends with a \instr{CALL} instruction to an external function, an edge is added from the node to the root. 

Given a bound $b$ on the number of function calls, we can systematically unfold the CFG so as to obtain all paths during which only $b$ or fewer functions are called. For instance, with a bound 2, the set of paths include all of those which visit $Node\_81\_87$ or $Node\_102\_109$ no more than twice. 

Statically constructing the CFG is non-trivial due to \emph{indirect jump}s in the bytecode generated by the Solidity compiler. For instance, part of bytecode for contract \emph{toyDAO} is shown as follows.

\vspace{-5pt}
{\scriptsize{\begin{verbatim}
  ...........                |       .......
   92 JUMPDEST               |       300 SHA3
   93 PUSH2 0x0064  //  100  |       301 DUP2
   96 PUSH2 0x0070  //  112  |       303 SSTORE
   99 JUMP                   |       304 POP
                             |
  100 JUMPDEST               |       305 JUMPDEST
  101 STOP                   |       306 POP
  .......                    |       307 JUMP
  112 JUMPDEST               |       ........
  113 PUSH1 0x00             |
  115 PUSH1 0x14             |
  .......                    |
\end{verbatim}}}
\vspace{-5pt}

\vspace{-5pt}

Considering that Solidity compiler use templates and often introduces indirect jumps (e.g., \instr{PUSH}), we actually construct CFGs from EVM opcode as follows:
\begin{compactitem}
  \item Disassemble the bytecode to a sequence of opcode instructions.
  \item Identify all basic blocks (BBL) from the opcode instructions as nodes of a CFG, where the boundaries among BBLs are branching instructions \instr{JUMP} and \instr{JUMPI}, \instr{JUMPDEST}, call instructions \instr{CALL}, and terminal instructions such as \instr{RETURN}, \instr{STOP}, and \instr{REVERT}.)
  \item Connect basic blocks with edges (e.g., direct jumps) which are statically decided from the opcode instructions.
  \item Use stack simulation to complete the CFG with edges for indirect jumps.
\end{compactitem}

In the above, whenever there are indirect jumps, their targets 
cannot be decided by checking proceeding instructions and we have missing edges. These nodes are known as \emph{dangling blocks} and 
we introduce \emph{stack simulation} to find the successor of them. Stack simulation is similar to define-use analysis except that dangling blocks which are reachable from the entry BBL are processed first. That is, we find all paths from entry BBL to dangling blocks (e.g., the two paths from $Node\_0\_12$ to $Node\_305\_307$) and simulate instructions in each path following semantics of the instruction on the stack. Note that a dangling block ends with \instr{JUMP} may have multiple successors in the CFG. When we reach the \instr{JUMP} or \instr{JUMPI} in the dangling block, the content of the top stack entry shall be determined and we connect the dangling block with BBL which starts at the address as in the top stack entry. For instance, for dangling block $Node\_305\_307$, there is only one successor $Node\_100\_101$ in both paths which is pushed by the instruction at address $093$. We repeat above steps until all dangling blocks are processed.

\subsection{Identifying Monetary Paths}
Given a bound $b$ on the number of call depth (i.e., the number of function calls) and a bound on the loop iterations, there are still many paths in the CFG to be analyzed. For instance, there are $6$ paths in the \emph{toyDAO} contract with a call depth bound of 1 (and a loop bound of 5) and 1296 with a call depth bound of 4. This is known as the path explosion problem~\cite{anand2008demand}. 
In this work, we focus on money-related paths to avoid path explosion
as almost all vulnerabilities~\cite{atzei2017survey} are `money'-related.

A node is money-related if and only if its BBL contains any of following opcode instructions: \instr{CALL}, \instr{CREATE}, \instr{DELEGATECALL} or \instr{SELFDESTRUCT}. In general, one of these instructions must be used when Ether is transferred from one account to another. A path which traverses through a money-related node is considered money-related.
\footnote{Note that each opcode instruction in EVM is associated with some gas consumption which technically makes them money-related. Gas\cite{wood2014ethereum} is the cost of any transaction that can be utilized to measure actions on Ethereum platform. However, the gas consumption alone in most cases does not constitute vulnerabilities and therefore we do not consider them money-related. In Fig.~\ref{toyDAOCFG}, we visualize money-related nodes with black background (e.g., the node $Node\_112\_162$ with a \instr{CALL}  statement~\instr{msg.sender.call.value(value)()}).}

\subsection{Identifying Property-Violating Paths} \label{property}

Next, \newtool~prioritizes paths that violate critical properties. The objective is to prioritize those paths which may trigger violation of critical properties for user inspection. The properties are designed based on previously known vulnerabilities and they can be configured and extended in \newtool.

\textbf{\emph{Property: Respect the Limit.}} In \newtool, we allow users to set a limit on the amount of Ether transferred out of the contract's wallet. For each path, we statically check whether Ether is transferred out of the wallet and whether the transferred amount is potentially beyond the limit. To do so, for each path, we use a symbolic variable to simulate the remaining limit. Each time an amount is transferred out, we decrease the variable accordingly and check whether the remaining limit is less than zero. If so, the path potentially violates the property. Note that if we are unable to determine the exact amount to be transferred 
, we conservatively assume the limit may be broken.

\textbf{\emph{Property: Avoid Non-Existing Addresses.}}
Any hexadecimal string of length no greater than 40 is considered a valid (well-formed) address in Ethereum. If a non-existing address is used as the receiver of a transfer, the Solidity compiler does not generate any warning and the contract can be deployed on Ethereum successfully. If a transfer to a non-existing address is executed, Ethereum automatically registers a new address (after padding 0s in front of address so that its length becomes 160bits). Because this address is owned by nobody, no one can withdraw Ether in it since no one has the private key.

For every path which contains instruction \instr{CALL} or
\instr{SELFDESTRUCT}, \newtool~checks whether the address in the instruction
exists or not. This is done with the help of EtherScan 
Ethereum~\cite{Ethereum27:online}  (which can check whether an address is registered or not). 
A path which sends Ether to a non-existing address is considered to be violating the property. 
Currently, to minimize the number of requests to EtherScan, we only query external transactions, thus may lead to false positives when the address has only internal transactions. Of course, users can configure \newtool~to also check internal transactions.

\textbf{\emph{Property: Guard Suicide.}} \newtool~checks whether a path
would result in destructing the contract without constraints on the date or
block number, or the contract ownership. A contract may be designed to ``suicide'' (with the opcode \instr{SELFDESTRUCT}) after certain date or reaching certain number of blocks, and often by transferring the Ether in the contract wallet to the owner. 
A notorious example is Parity Wallet which resulted in an estimated loss of tokens worthy of \$155 million~\cite{AnotherP81:online}.

We thus check whether there exists a path which executes \instr{SELFDESTRUCT} and whether its path condition is constituted with constraints on date or block number and contract owner address. While checking the former is straightforward, checking the latter is achieved by checking whether the path contains constraints on instruction \instr{TIMESTAMP} or \instr{BLOCK}, and checking whether the path condition compares the variables representing the contract owner address with other addresses. A path which calls \instr{SELFDESTRUCT} without such constraints is considered a violation of the property.

\begin{figure}[t]
{\scriptsize
 \centering
  \begin{verbatim}
   contract StandardToken is Token {
1     function destroycontract(address _to) {
2        require(now > start + 10 days);
3        require(msg.sender != 0);
4        selfdestruct(_to);
5     }
6     ...
7  }
8  contract Problematic is StandardToken { ... }
    \end{verbatim}}
    \vspace{-15pt}
 \caption{Guardless suicide}
 \label{contract:GuardlessSuicide}
 \vspace{-15pt}
\end{figure}

One example is the \emph{Problematic} contract\footnote{We hide the names of the contracts as some of them are yet to be fixed.} shown in Fig.~\ref{contract:GuardlessSuicide}. Contract \emph{Problematic} inherits contract \emph{StandardToken}, 
where one of functions is \emph{destroycontract()} allowing one to destruct contract. \newtool~can report that line 4 potentially violates the property.

\textbf{\emph{Property: Be No Black Hole.}} In a few cases,
\newtool~analyzes paths which do not contain \instr{CALL},
\instr{CREATE}, \instr{DELEGATECALL} or \instr{SELFDESTRUCT}. For instance, if a contract has no money-related paths (i.e., never sends any Ether out), \newtool~then checks whether there exists a path which allows contract to receive Ether. The idea is to check whether contract acts like a black hole for Ether. If it does, it is considered a vulnerability.

To check whether the contract can receive Ether, we check whether there is
a \emph{payable} function. Since Solidity version 0.4.x, a contract is allowed
to receive Ether only if one of its public functions is declared with the
keyword \emph{payable}. When the Solidity compiler compiles a non-payable
function, the following sequence of opcode instructions are inserted before the function body.

\begin{infobox}{\scriptsize\begin{verbatim}
    1  CALLVALUE
    2  ISZERO
    3  PUSH XX
    4  JUMPI
    5  PUSH1 0x00
    6  DUP1
    7  REVERT
\end{verbatim}}\end{infobox}

At line 1, the instruction \instr{CALLVALUE} retrieves the message
value (to be received). Instruction \instr{ISZERO} then checks if the value is
zero, if it is zero, it jumps (through the \instr{JUMPI} instruction at line 4) to the
address which is pushed into stack by the instruction at line 3; or it goes to
the block starting at line 5, which reverts the transaction (by instruction
\instr{REVERT} at line 7). Thus, to check whether the contract is allowed to receive Ether, we go through every path to check whether it contains the above-mentioned sequence of instructions. If all of them do, we conclude that the contract is not allowed to receive Ether. Otherwise, it is.

If the contract can receive Ether but cannot send any out, we identify the path for receiving Ether as potentially violating the property and label it with a warning messaging stating that the contract is a black hole.

Above properties are designed based on reported vulnerabilities. Of course, \newtool~is designed to be extensible, i.e., new properties can be easily supported by providing a function which takes a path as input and reports whether the property is violated.

To further help users understand paths of a smart contract, \newtool~supports additional analysis. For instance, \newtool~provides analysis of gas consumption of paths. 

However, without trying out all possible inputs, users may not be aware of the existence of certain particularly gas consuming paths. 
The gas consumption of a path is estimated based on each opcode instruction in the path statically.

\subsection{Ranking Program Paths} \label{Prioritizing}
To allow user to focus on most critical paths and to save analyses efforts, we prioritize paths according to the likelihood they reveal critical vulnerability. For each path, we calculate a criticalness score and rank paths according to scores. Criticalness scores are calculated as follows: let $pa$ be a path and $V$ be the set of properties which $pa$ violates.
\begin{equation}
criticalness(pa) = \frac{\Sigma_{pr \in V} \alpha_{pr}}{\epsilon * bound(pa)}
\end{equation}
where $\alpha_{pr}$ is a constant which denotes the criticalness of violating property $pr$, $bound(pa)$ is the depth bound of path $pa$ (i.e., the number of function calls) and $\epsilon$ is a positive constant. Intuitively, the criticalness is designed such that the more critical a property the path violates, the larger the score is; and the more properties it violates, the larger the score is. Furthermore, it penalizes long paths so that short paths are presented first for user inspection.

\begin{table}[t]
\centering
    \footnotesize
    \setlength\tabcolsep{3pt}
\caption{Definition of $\alpha_{pr}$}
\label{table:pr-definition}
\begin{tabular}{ccccc}
\hline
 & transfer limit & non-existing addr. & suicide & black hole  \\ \hline
Likelihood & 1 & 1 & 2 & 3 \\ 
Severity & 2 & 3 & 3 & 2 \\ 
Difficulty & 2 & 2 & 3 & 2 \\ 
$\alpha_{pr}$ & 4 & 6 & 18 & 12 \\ \hline
\end{tabular}
\vspace{-8pt}
\end{table}

To assess the criticalness of each property, we use the technique called
failure mode and effects analysis (FMEA~\cite{stamatis2003fmea}) which is a risk management tool widely used in a variety of industries. FMEA evaluates each property with 3 factors, i.e., \emph{Likelihood}, \emph{Severity} and \emph{Difficulty}. Each factor is a value rating from 1 to 3, i.e., 3 for \emph{Likelihood} means the most likely; 3 for \emph{Severity} means the most severe and 3 for \emph{Difficulty} means the most difficult to detect. The criticalness $\alpha_{pr}$ is then set as the product of the three factors. After ranking
, only paths which have a criticalness score larger than certain threshold are subject to further analysis, reducing the number of paths significantly.

In order to identify the threshold for criticalness, we adapt the k-fold cross-validation\cite{devijver1982pattern,kohavi1995study} idea in statistical area. We collected a large set of smart contracts and split them into a training data set(10,452 contracts) and a test data set (25,678 contracts). 
We repeated the experiments 20 times which took more than 5,700 total hours of all machines and optimizes those parameters.
The adapted parameters are shown in Table~\ref{table:pr-definition}, and $\epsilon$ is set to be 1 and the threshold for criticalness is set to be 10.

\subsection{Feasibility Checking}
Not all the paths are feasible. To avoid such false alarms, we filter infeasible paths through symbolic execution~\cite{king1976symbolic}.
The basic idea is to symbolically execute a given program.
Symbolic execution has been previously applied to Solidity programs in Oyente~\cite{luu2016making} and MAIAN~\cite{nikolic2018finding}. In this work, we apply symbolic execution to reduce the paths which are to be presented for users' inspection. Only if a path is found to be infeasible by symbolic execution, we remove it. In comparison, both Oyente and MAIAN aim to fully automatically analyze smart contracts and thus when a path cannot be determined by symbolic execution, the result may be a false positive or negative.

\begin{figure}[!h]
{\scriptsize
  \begin{verbatim}
  contract GigsToken {
1   function createTokens() payable {
2     require(msg.value > 0);
3     uint256 tokens = msg.value.mul(RATE);
4     balances[msg.sender] = balances[msg.sender].add(tokens);
5     owner.transfer(msg.value);
6   }
7   ...
  }
  \end{verbatim}}
  \vspace{-10pt}
 \caption{A non-greedy contract}
 \label{contract:FalseGreedy}

\end{figure}

For instance, Fig.~\ref{contract:FalseGreedy} shows a contract which is capable of receiving (since the function is \emph{payable}) and sending Ether (due to \emph{owner.transfer(msg.value)} at line 5), and thus \newtool~does not flag it to be a black hole contract. MAIAN however claims that it is. A closer investigation reveals that because MAIAN has trouble in solving path conditions for reaching line 5, and thus mistakenly assumes the path is infeasible. As a result, it believes there is no way Ethers can be sent out and thus the contract is a black hole.

\section{Implementation and Evaluation}
\label{experiment}

\subsection{Implementation}
\newtool~is implemented in C++ with about 8K lines of code. 
The symbolic execution engine in \newtool~is built based on the Z3 SMT solver~\cite{de2008z3}. 

\subsection{Experiment} %
\label{sub:experiment}
We aim to to answer research questions (RQ) regarding \newtool's efficiency, effectiveness and usefulness in practice. Our test subjects contain all 36,099 contracts (including both the training set and the test set) with Solidity source code downloaded from EtherScan. \newtool~can directly take EVM code as input and the source code is used for our manual inspection for experiment purpose.

All experiment are done on an Amazon EC2 C3 xlarge instance installed with Ubuntu 16.04 and gcc 5.4. The timeout set for \newtool~is: global wall time is 60 seconds and Z3 solver timeout is 100 milliseconds. The limit on the maximum number of blocks for a single path is set to be 60, and the limit on the maximum iterations of loops is set to be 5, i.e., each loop is unfolded at most five times.

\textbf{\emph{RQ1: Is \newtool~efficient enough for practical usage?}}
In this experiment, we evaluate \newtool~in terms of its execution time. We systematically apply \newtool~to all the benchmark programs in the training set.

\begin{figure}[t]
    \begin{minipage}{.5\columnwidth}
        \includegraphics[width=.9\columnwidth,scale=0.8]{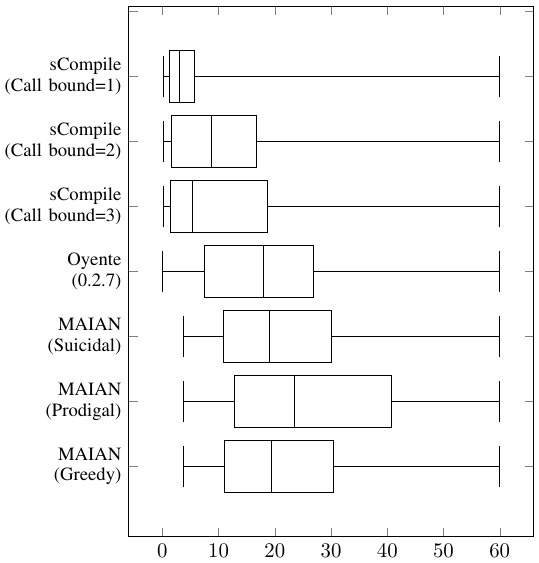}
       
    \end{minipage}
    \begin{minipage}{.4\columnwidth }
         \resizebox{.8\columnwidth}{!}{%
        \begin{tabular}{|l|l|l|}
        \hline
         & median(s) & timeout \# \\ \hline
        \begin{tabular}[c]{@{}l@{}}sCompile\\ (Call bound = 1)\end{tabular} & 3.106 & 1145 \\ \hline
        \begin{tabular}[c]{@{}l@{}}sCompile\\ (Call bound = 2)\end{tabular} & 8.717 & 1737 \\ \hline
        \begin{tabular}[c]{@{}l@{}}sCompile\\ (Call bound = 3)\end{tabular} & 5.267 & 2597 \\ \hline
        Oyente & 18.015 & 2223 \\ \hline
        \begin{tabular}[c]{@{}l@{}}MAIAN\\ (Suicidal)\end{tabular} & 19.053 & 1561 \\ \hline
        \begin{tabular}[c]{@{}l@{}}MAIAN\\ (Prodigal)\end{tabular} & 23.472 & 6186 \\ \hline
        \begin{tabular}[c]{@{}l@{}}MAIAN\\ (Greedy)\end{tabular} & 19.397 & 1081 \\ \hline
        \end{tabular}
        }
        \label{thelabeltab}
    \end{minipage}%
    \caption{Execution time of~\newtoolBold~vs. Oyente vs. MAIAN}
    \label{exe_time}
\end{figure}

The results are summarized in Figure ~\ref{exe_time}. In sub-table of Figure~\ref{exe_time}, the second, third and
fourth row show the execution of \newtool~with call depth bound 1, 2, and
3 respectively.
For comparison, the fifth row shows the execution time of Oyente
(the latest version 0.2.7) with the same timeout. We remark that the comparison
should be taken with a grain of salt. Oyente does not consider sequences of
function calls, i.e., its bound on function calls is 1. Furthermore, it does
not consider initialization of variables in the constructor (or in the contract
itself). The next columns show the execution time of MAIAN (the latest commit
version on Mar 19). Although MAIAN is designed to analyze paths with
multiple (by default, 3) function calls, it does not consider the possibility of a third-party contract calling any function in
the contract through inter-contract function calls and thus often explores much fewer paths
than~\newtool. Furthermore, MAIAN checks only one of the three properties
(i.e., suicidal, prodigal and greedy) each time. Thus, we must run MAIAN three
times to check all three properties. The different bounds used in all three tools are summarized in Table~\ref{exe_bound_table1}.

\begin{table}[t]
\centering
    \footnotesize
    \setlength\tabcolsep{3pt}
\caption{Loop bound definitions among three tools}
\label{exe_bound_table1}
\vspace{-5pt}
\resizebox{.6\columnwidth}{!}{%
\begin{tabular}{ccccc}
\hline
Tool & call bound & loop bound & timeout & other bound \\ \hline
\newtoolBold & 3 & 5 & 60 s & 60 cfg nodes \\ 
\textbf{Oyente} & 1 & 10 & 60 s & N.A. \\ 
\textbf{MAIAN} & 3 (no inter-contract) & N.A. & 60 s & 60 cfg\ nodes \\ \hline
\end{tabular}
}
\vspace{-15pt}
\end{table}

In sub-table of Figure~\ref{exe_time}, the second column shows the median execution time and
the third column shows the number of times the execution time exceeds the global
wall time ($60$ seconds). We observe that \newtool~almost always finishes its
analysis within $10$ second. Furthermore, the execution time remains similar
with different call depth bounds. This is largely due to \newtool's strategy on
applying symbolic execution only to a small number of top ranked critical
paths. We do however observe that the number of timeouts increases with
an increased call depth bound. A close investigation shows that this is mainly
because the number of paths extracted from CFG is much larger and it takes
more time to extract all paths for ranking. In comparison, although Oyente
has a call depth bound of 1, it times out on more contracts and spends more
time on average. MAIAN spends more time on each property than the total execution of
\newtool. For some property (such as \emph{Greedy}), MAIAN times out fewer
times, which is mainly because it does not consider inter-contract function
calls and thus works with a smaller CFG.

The sub-figure in Figure ~\ref{exe_time} visualizes the distribution of execution time of the tools in plot-box. The x-axis represents the execution time (in seconds). 
From the figure, we can conclude that \newtool~is efficient.

Table~\ref{numbers} shows the statistics on the number of processed paths, including the estimated total number of paths on average (in the second column), the number of symbolic-executed (based on CFG), and the number passed to users. 
It can be observed that only a small fraction of the paths
are symbolically analyzed. Furthermore, the number of symbolically executed
paths remain small even when the call depth bound is increased. This is because only the top ranked critical paths are analyzed by symbolic execution.

\begin{table}[t]
\centering
    \footnotesize
    \setlength\tabcolsep{3pt}
\caption{Average number of program paths}
\label{numbers}
\vspace{-3pt}
\begin{tabular}{cccc}
\hline
& \textbf{In total} & \textbf{Symbolic-executed} & \textbf{To user} \\
\hline
call depth 1 &48.92 &37.51 & 1.49\\ 
call depth 2 &6177.21 &144.24 & 12.46\\ 
call depth 3 &31346.62 &121.23 & 12.62\\ \hline
\end{tabular}
\vspace{-8pt}
\end{table}

\begin{table}[t]
\centering
    \footnotesize
    \setlength\tabcolsep{3pt}
\caption{Comparison on vulnerable contracts}
\label{table:comparison on property2&3&4}
\vspace{-3pt}
\resizebox{\columnwidth}{!}{%
\begin{tabular}{c|ccc|ccc}
\hline
 & \multicolumn{3}{c|}{\newtoolBold}  & \multicolumn{3}{c}{\textbf{MAIAN}} \\ 
 & alarmed & true positive & false positive & alarmed & true positive & false positive \\ \hline
\emph{Avoid non-existing address} & 37 & 32 & 5 & N.A. &N.A. & N.A. \\ 
\emph{Be no black hole}  & 57 & 57 & 0 & 141 & 56 & 85 \\ 
\emph{Guard suicide} & 42 & 38 & 4 & 66 & 30 & 36 \\ \hline
\end{tabular}
}
\vspace{-8pt}
\end{table}
\textbf{}

\textbf{\emph{RQ2: Is~\newtool~effective to practical usage?}} In the second experiment, we aim to investigate the effectiveness of~\newtool.
We apply \newtool~to all 36,099 contracts and manually inspect the critical paths reported by \newtool~to check whether the path, together with the associated warning message, reveals a true vulnerability in the contract. Note that not all properties checked by \newtool~readily signals a vulnerability. 
We only focus on those results produced by \newtool~which are directly related to vulnerabilities in the following, i.e., paths which are deemed to violate property ``avoid non-existing addresses'', ``be no black hole'' and ``guard suicide''. Note that two of the properties (i.e., the latter two) analyzed by \newtool~are supported by MAIAN as well. We can thus compare \newtool's performance with that of MAIAN for these two properties. The results are shown in Table~\ref{table:comparison on property2&3&4}. In the following, we discuss the detailed findings\footnote{We have informed all developers whose contact info are available about the vulnerabilities in their contracts and several have confirmed the vulnerabilities and deployed new contracts to substitute the vulnerable ones. Some are yet to respond, although the balance in their contracts are typically small.}.  

For \emph{Property: Be no Black Hole}, there are 57 contracts in the
training set are marked vulnerable by \newtool. We manually confirmed that they are all true positives. In comparison, MAIAN identified 141 black hole contracts and 56 contracts among them are true positives, 43 of which overlap with \newtool's results. 
For 13 missed contracts by \newtool~but detected by MAIAN, 
all of them took more than 60 seconds and thus \newtool~timed out before finishing analyzing. 

The other 85 identified by MAIAN are false positives and 
62 of them are library contracts. 
We randomly choose 5 contracts from the remaining for further
investigation. We find Z3 could not finish solving the path condition in time
and thus MAIAN conservatively marks the contract as vulnerable. After extending
the time limit for Z3 and total timeout, 4 of the 5 false positives are
still reported. The reason is that these contracts can only send Ether out
after certain period, and MAIAN could not find a feasible path to send Ether
out for such cases, and mistakenly flags contract as a black hole. 

For \emph{Property: Guard Suicide}, \newtool~reports a program path
if it leads to \instr{SELFDESTRUCT}, without a constraint on the ownership of the contract or the date or the block number, i.e., a guard to prevent an unauthorized users from killing the contract. Among the analyzed contracts, \newtool~identified 42 contracts which contain at least one path which violates the property. 
Many of the identified contracts
violate the property due to contract inheritance as shown in
Fig.~\ref{contract:GuardlessSuicide}. 

The remaining 4 cases reported by~\newtool~are false positives. We manually
investigated into them and found that they belong to two uncommon coding cases (where 3 of them are originated from the same contract) and three of them can be detected by \newtool~by slightly revising its implementation.

MAIAN identified 66 contracts violating the property. 30 of them are true positives, 13 of which are also identified by \newtool. The other 36 are false positives. The contract \emph{MiCarsToken} shown in Fig.~\ref{contract:AmbiguousCase} shows a typical false alarm. There are 2 constraints before \instr{SELFDESTRUCT} in the contract. \newtool~considers such a contract safe for there is a guard of \emph{msg.sender == owner} (or the other condition), whereas MAIAN reports a vulnerability as the contract can also be killed if the msg.sender is not the owner when the second condition is satisfied.

\begin{figure}[t]
{\scriptsize 
  \begin{verbatim}
  contract MiCarsToken {
      function killContract () payable external {
        if (msg.sender==owner ||
                msg.value >=howManyEtherInWeiToKillContract)
            selfdestruct(owner);
      }
      ...
  }
  \end{verbatim}}
  \vspace{-10pt}
 \caption{Ambiguous cases between \newtool~and MAIAN }
 \label{contract:AmbiguousCase}
\end{figure}

We further analyzed the $17$ cases which were neglected by \newtool. 6 of them are alarmed for owner change as exemplified in Fig.~\ref{contract:ownerChange}. In this contract, \emph{selfdestruct} is well guarded, but the developer makes a mistake so that the constructor becomes a normal function, and anyone can invoke \emph{mortal()} to make himself the owner of this contract and kill the contract.

\begin{figure}[t]
{\scriptsize \vspace{-5pt}
  \begin{verbatim}
  contract Mortal {
      address public owner;
      function mortal() { owner = msg.sender; }
      function kill() {
        if (msg.sender == owner) suicide(owner); }
  }
  \end{verbatim}}
  \vspace{-5pt}
 \caption{Contract of owner change}
 \label{contract:ownerChange}
 \vspace{-12pt}
\end{figure}

For \emph{Property: Avoid Non-existing Address}
. For the contracts in the training set, all addresses identified are of length 160 bits. However, there are 37 contracts identified as non-existing addresses (i.e., not registered in Ethereum mainnet). They may be used for different reasons. For example, in contract \emph{AmbrosusSale}, the address of TREASURY does not exist before the function \texttt{specialPurchase()} or \texttt{processPurchase()} is invoked (which will cost more gas for its first user). 
And there are 5 addresses registered by internal transactions. 

We further analyzed 25,647 contracts newly uploaded in EtherScan from
February 2018 to July 2018. For \emph{``Be no Black Hole''}, there are 109
vulnerabilities out of 139 alarms generated by \newtool. Applying MAIAN on
these contracts, 84 of them are marked vulnerable, 77 of which are true
vulnerabilities overlapping with those found by \newtool~and 7 library contracts are marked vulnerable mistakenly. Among the 139 contracts, 25 vulnerable ones are missed by MAIAN according to our manual check. For \emph{``Guard Suicide''}, there are 83 vulnerabilities out of 114 alarms generated by \newtool. Applying MAIAN on these contracts, 42 are marked vulnerable, all of which overlap with those found by \newtool.
For \emph{``Avoid Non-existing Addresses''}, there are 80 vulnerabilities out of 87 alarms generated by \newtool. The 7 false alarms are due to internal transactions. 

In total, \newtool~identifies 224 vulnerabilities from the 36,099 contracts consisting of 46 \emph{Black Hole} vulnerabilities, 66 \emph{Guardless Suicide} vulnerabilities and 112 \emph{Non-existing Address} vulnerabilities.

\textbf{\emph{RQ3: Is~\newtool~useful to contract users?}} Different from
other tools which aim to fully automatically analyze smart contracts, \newtool~is
designed to facilitate human users. We thus conduct a user study to see whether \newtool~is helpful to them.

The study takes the form of an online test. Once a user starts the test, first the user is briefed with necessary background on smart contract vulnerabilities (with examples). Then, 6 smart contracts (selected at random each time from a pool of contracts) are displayed one by one. For each contract, the source code is first shown. Afterwards, the user is asked to analyze the contract and answer the two questions. The first question asks what is the vulnerability the contract has. The second question requires user to identify the most gas consuming path in contract (with one function call).

For the first three contracts, the outputs from \newtool~are shown alongside the contract source code as a hint to the user. For the remaining 3 contracts,
the hints are not shown. The contracts are randomized so that not the same contracts are always displayed with the hint. The goal is to check whether users can identify the vulnerabilities correctly and more efficiently with \newtool's results.

We distribute the test through social networks and online professional forums.
We also distribute it through personal contacts who we know have some experience
with Solidity smart contracts. In three weeks we collected 48 successful
responses to the contracts (without junk
answers)\footnote{There are about 80 people who tried the test. Most of the respondents however leave the test after the first question, which perhaps evidences the difficulty in analyzing smart contracts.}.
Table~\ref{table:survey} summarizes the results. Recall that \newtool's results are presented for the first three contracts. Column LOC and \#paths shows
the number of lines and paths in each contract. Note that in order to
keep the test manageable, we are limited to relatively small contracts in this study. Columns Q1 and Q2 show the number of correct
responses (the numerator) out of the number of valid responses (the
denominator). We collect the time (in seconds) taken by each user in the Time
column to answer all the questions. In the end of the survey we ask the user to
give us a score (on the scale of 1 to 7, the higher the score the more useful
our tool is) on how useful the hints in helping them answer the questions. The
value in column Usefulness is the average score over all responses because all
responses are shown half the hints.

\begin{table}[t]
\centering
    \footnotesize
    \setlength\tabcolsep{3pt}
\caption{Statistics and results of surveyed contracts}
\vspace{-8pt}
\label{table:survey}
\begin{tabular}{c | c c | c c| c| c}
\hline
Contract & LOC & \#paths & Q1 & Q2 & Time & Usefulness\\
\hline
 C1 (w)  & 33 & 8 & 7/8 & 3/8 & 119 & \multirow{6}{*}{5}\\
 C2 (w) & 52 & 16 & 7/8 & 2/8 & 98 & \\
 C3 (w) & 67 & 38 & 7/8 & 2/8 & 233 & \\
 C4 (w/o) & 87 & 59 & 2/8 & 1/8 & 414 & \\
 C5 (w/o) & 103 & 13 & 3/8 & 1/8 & 397 & \\
 C6 (w/o) & 107 & 27 & 4/8 & 1/8  & 420 & \\
\hline
\end{tabular}
\vspace{-8pt}
\end{table}

The results show that for the first three contracts for which \newtool's analysis results are shown, almost all users
are able to answer Q1 correctly using less time. For the last three
contracts without the hints, most of the users cannot identify the vulnerability correctly and it takes more
time for them to answer the question. For identifying the most gas-consuming path, even with the hints on which
function takes the most gas, most of the users find it difficult in answering the question, although with \newtool's help, more users are able to answer the question correctly. The results show that gas consumption is not a well-understood
problem and highlight the necessity of reporting the condition under which
maximum gas consumption happens. All the users think our tool is
useful (average score is $5/7$) in helping them identify the problems.

\section{Related work} \label{related}
\newtool~is related to existing work on identifying vulnerabilities in smart contracts that can be roughly categorized into 3 groups according to the level
at which the vulnerability resides at: Solidity-level, EVM-level, and
blockchain-level~\cite{atzei2017survey}. In addition, existing work can be
categorized according to the techniques they employ to find vulnerabilities:
symbolic execution~\cite{luu2016making,nikolic2018finding,mythril,manticore,jiang2018contractfuzzer},
static-analysis based approaches~\cite{securify} and formal verification~\cite{kalra2018zeus,Fstar}. Our approach works at the EVM-level and is based on static analysis and symbolic execution, and is thus closely related to the following work.

Oyente~\cite{luu2016making} formulates the security bugs as intra-procedural properties and uses symbolic execution to check these properties. 
However, Oyente does not perform inter-procedural analyses to check inter-procedural or trace properties as did in \newtool.

MAIAN~\cite{nikolic2018finding} is recently developed to find three types of
problematic contracts in the wild: prodigal, greedy and suicidal. It formulates
the three types of problems as inter-procedural properties and performs bounded
inter-procedural symbolic execution. It builds a private testnet to valid
whether the contracts found by it are true positives by executing the contracts
with data generated by symbolic execution. However, \newtool~differs from MAIAN in following
aspects. First, \newtool~makes a much more conservative assumption about a call
to third-party contract which we assume can call back a function in
current contract. \newtool~is designed to reduce user effort rather than to analyze smart contracts fully automatically. Secondly, \newtool~supports more properties than MAIAN. Thirdly, \newtool~checks properties in ways which are different from MAIAN. 
Other symbolic execution based tools~\cite{mythril,manticore} perform
intra-procedural symbolic analysis directly on the EVM bytecode as what Oyente
does.

The tool Securify~\cite{securify} is based on static analysis to analyze
contracts. It specifies both compliance and violation patterns for the property. The
vulnerability detection problem is then reduced to search the patterns on the
inferred data and control dependencies information. The use of compliance
pattern reduces the number of false positives in the reported warnings. In the
ranking algorithm, our approach rely on syntactic information to reduce paths for further
symbolic analysis to improve performance. We analyze the extracted paths with
symbolic execution which is more precise than the pure static analysis as adopted by Securify.

Other attempts on analyzing smart contracts include formal verification using either model-checking techniques~\cite{kalra2018zeus} or theorem-proving approaches~\cite{Fstar}. They in theory can check arbitrary properties specified manually in a form accepted by the model checker or the theorem prover. It is known that model checking has limited scalability whereas theorem proving requires an overwhelming amount of user effort.

\section{Conclusion} \label{conclusion}
We proposed a practical approach named \newtool~to reveal ``money-related`` paths in smart contract and to further detect vulnerabilities among critical ones. In our experiment among 36,099 smart contracts, it detected $224$ new vulnerabilities. All the new vulnerabilities are well defined in our approach and could be presented to the user in well-organized information within a reasonable time frame. A comparison with two existing approaches also demonstrated that \newtool~is both efficient and effective. \\

\noindent\textbf{Acknowledgement}. This work is supported by the Singapore Ministry of Education (MOE) Academic Research Fund (AcRF) Tier 1 grant, the Youth Innovation Promotion Association of the Chinese Academy of Sciences (YICAS) (Grant No. 2017151), the Young Elite Scientists Sponsorship Program by CAST (Grant No. 2017QNRC001), and the Blockchain Technology and Application Joint Laboratory, Guiyang Academy of Information Technology (Institute of Software Chinese Academy of Sciences Guiyang Branch).

\bibliographystyle{unsrt}
\bibliographystyle{splncs04}
\bibliography{scompile}

\end{document}